\documentclass[useAMS]{mn2e}
\usepackage{graphicx}
\def\ltsima{$\; \buildrel < \over \sim \;$}
\def\gtsima{$\; \buildrel > \over \sim \;$}
\def\lsim{\lower.5ex\hbox{\ltsima}}
\def\gsim{\lower.5ex\hbox{\gtsima}}

\title[The peculiar Horizontal Branch of NGC~2808.]
   {The peculiar Horizontal Branch of NGC~2808.}

\author[E. Dalessandro et al.]
 {E.~Dalessandro,$^1$ M.~Salaris,$^2$ F.R.~Ferraro,$^1$
  S.~Cassisi,$^3$ B.~Lanzoni,$^1$ R.T.~Rood,$^4$ 
  \newauthor F.~Fusi Pecci,$^5$ E.~Sabbi$^6$ \\
  $^1$Dipartimento di Astronomia, Universit\`a degli Studi
 di Bologna, via Ranzani 1, I--40127 Bologna, Italy\\
 $^2$Astrophysics Research Institute, Liverpool John Moores University,
 Twelve Quays House, Egerton Wharf, Birkenhead CH41 1LD, UK\\  
 $^3$INAF - Osservatorio Astronomico di Collurania, via Mentore Maggini, 
 64100 Teramo, Italy\\
 $^4$Astronomy Department, University of Virginia,
 P.O. Box 400325, Charlottesville, VA, 22904, USA\\
 $^5$INAF - Osservatorio Astronomico di Bologna, via Ranzani 1, 
 40127 Bologna, Italy\\
 $^6$Space Telescope Science Institute 3700 San Martin Drive, Baltimore,
    MD, 21218, USA}
\date{3 Aug, 2010}

\pagerange{\pageref{firstpage}--\pageref{lastpage}} \pubyear{2010}

\def\LaTeX{L\kern-.36em\raise.3ex\hbox{a}\kern-.15em
    T\kern-.1667em\lower.7ex\hbox{E}\kern-.125emX}

\begin{document} 

\label{firstpage}

\maketitle
 
\begin{abstract}
We present an accurate analysis of the peculiar Horizontal Branch (HB)
of the massive Galactic globular cluster NGC~2808, based on
high-resolution far-UV and optical images of the central region of the
cluster obtained with HST.  We confirm the multimodal distribution of
stars along the HB: 4 sub-populations separated by gaps are
distinguishable.  The detailed comparison with suitable theoretical
models showed that {\it(i)} it is not possible to reproduce the
luminosity of the entire HB with a single helium abundance, while an
appropriate modeling is possible for three HB groups by assuming
different helium abundances in the range
$0.24<\Delta Y<0.4$ that are 
consistent with the multiple populations observed in the Main
Sequence; {\it(ii)} canonical HB models are not able to properly match
the observational properties of the stars populating the hottest end
of the observed HB distribution, the so called ``blue hook
region.'' These objects are probably ``hot-flashers,'' stars that peel
off the red giant branch before reaching the tip and ignite helium at
high $T_{\rm eff}$.  Both of these conclusions are based on the luminosity
of the HB in the optical and UV bands and do not depend
on specific assumptions about mass loss.
\end{abstract} 


\section{INTRODUCTION}

Galactic globular clusters (GCs) are very old (age $ \sim 10$--13 Gyr)
stellar systems populated by $10^5$--$ 10^6$ stars. Initially GCs were
thought to be prime examples of simple stellar populations, i.e.,
populations of coeval stars born with the same initial chemical
composition.  That situation began to change in the late 70's with the
first spectroscopic surveys that disclosed star-to-star chemical
variations of light element abundances within individual clusters
(e.g., Cohen~1978).  These variations are now known to be present in a
large fraction of GCs and appear in the form of well defined
anticorrelations between the abundances of C-N, Na-O and Mg-Al pairs,
that cannot be explained in terms of evolutionary/mixing effects (see,
e.g., Gratton et al.~2000, Carretta et al.~2008, 2009).  In addition
there is growing observational evidence for helium (He) abundance
variations within a few clusters which we discuss below. The presence
of specific chemical patterns supports a scenario invoking multiple
star formation events in some, perhaps most, GCs.  These took place at
the very beginning of the cluster evolution on relatively short
time-scales, of the order of $100\,$Myr.  The younger generations
could have been born out of gas enriched by the winds of intermediate
mass Asymptotic Giant Branch stars (AGBs; e.g., Ventura et al.~2009)
or massive fast rotating stars (Decressin et al. 2007) formed during
the first star formation episode. Dynamical simulations, e.g., by
D'Ercole et al.~(2008) and Decressin et al. (2008), have investigated
how these winds could have been retained by the clusters an
incorporated into future generations of stars.

One of the challenges facing self-enrichment scenarios in a few
clusters is that they must be able to account for appreciable
differences in the helium mass fraction ($Y$) between different
generations, with $Y$ reaching values up to twice the primordial
abundance while at the same time not appreciably altering the
abundances of most other elements like iron.  A notable exception is
$\omega$ Centauri (NGC~5139), where the Main Sequence (MS) splitting
detected by Bedin et al. 2004 has been interpreted as due to He
variation (with the richest population having $Y\sim0.38$; Piotto et
al.~2005) and where at least five distinct populations with different
iron content and possibly different ages have been revealed (Pancino
et al.~2002, Ferraro et al.~2004, Sollima et al.~2005, Villanova et
al.~2007, Calamida et al.~2009).  However $\omega\,$Cen may not
be a genuine GC, but rather the remnant of a dwarf
galaxy partially disrupted by the interaction with the Milky Way
(Bekki \& Freeman~2003, Mackey \& Van den Bergh~2005). A similar
system (Terzan~5) harboring two distinct populations with different
iron content and (possibly) ages has been recently found in the
Galactic Bulge (Ferraro et al.~2009). Similarly, this system is
suspected to be not a genuine GC, but instead the remnant of a larger
system that contributed to the formation of the Galactic Bulge. A
milder evidence of internal spread of the metal content ($\Delta$
[Fe/H] $\sim 0.1$) has been observed also in M22 (Marino et al. 2009;
Da Costa et al. 2009).

In the context of the multi-population scenarios, one of the most
puzzling cases is that of NGC~2808. 
The first photometric observations (Harris~1974, Ferraro et
al.~1990) revealed a Horizontal Branch (HB) with a very complex
structure.  The cluster HB is well populated both at colours redder
than the RR~Lyrae instability strip and along its hot blue tail (BT)
that covers a range of $\sim5$ mag below the mean level of the
instability strip.  This morphology is not easily explained in terms of
the cluster metallicity (${\rm [Fe/H]}\sim-1.3$), since it is well
outside the common paradigm that links red HBs to metal-rich GCs and
blue-HBs to metal-poor systems.  In this sense NGC~2808 is similar to
other very massive (and much more metal rich) GCs, like NGC~6388 and
NGC~6441 (Rich et al.~1997, Busso et
al.~2007, Dalessandro et al.~2008).  The cluster colour-magnitude-diagram (CMD)  also shows puzzling
discontinuities in the
stellar distribution along the BT (Sosin et al.~1997, also Bedin et
al.~2000, Castellani et al.~2006, Iannicola et al.~2009) similar to those found in several
other GCs (Ferraro et al.~1998).

D'Antona et al.~(2005) first noticed a broadening of the NGC~2808 MS
which was incompatible with photometric errors. They suggested that
the MS consisted of two components with the same age and [Fe/H], but
different initial He abundances, with $Y\sim0.4$ for the He-rich
component. The complex MS structure was confirmed by an accurate
photometric and proper motion analysis with deep Hubble Space
Telescope (HST) data (Piotto et al.~2007). They found that the MS of
NGC~2808 splits into three sub-populations, all with age
$\sim12.5$\,Gyr and $Y\sim0.248$ for the red-MS, $Y\sim0.30$ for the
mean-MS and $0.35<Y<0.40$ for the blue-MS. D'Antona et al. (2005)
further hypothesized that the complex HB is connected to the MS
components with different He abundances. \footnote{Even though
NGC~2808 is one of the most massive clusters and shows the presence
of three distinct stellar populations characterized by quite different
He abundances, ther is no evidence of appreciable [Fe/H] variations
(Carretta et al. 2006).} However their HB morphology analysis was
performed in the optical bands where the increase of the
bolometric corrections at high effective temperature turns the HB into
an almost vertical structure (rather than horizontal) when $T>10,000\,
$K.  Along this vertical part of the HB, models with different initial
He abundance tend to overlap, and the identification of
sub-populations with different initial $Y$ is not independent of
assumptions about the amount of mass lost along the RGB phase.
Ferraro et al. (1998) and Rood et al. (2008) suggest that the optimal
diagram for the study of blue HB stars is the ($m_{F160BW}$,
$m_{F160BW}-m_{F555W}$) CMD in the HST filter system.  In this CMD the
hottest HB stars are the most luminous and lie along almost horizontal
sequences, whose luminosity is very sensitive to the initial $Y$
abundance irrespective of the precise value of the stellar mass.

Here we present an accurate photometric study of the central regions
of NGC~2808, based on high-resolution HST observations in
far-UV and optical filters. With evolutionary
models and constraints on the initial $Y$ distribution coming from the
analysis of the multiple MS, we investigate the complex structure of
the HB of this massive GC.
The paper is structured as follows. Section~2 describes the
observational data, their reduction and calibration, Sect.~3 presents
an analysis of the observed properties of the cluster HB, and Sect.~4
details their interpretation in term of theoretical models. A summary
and the conclusions follow in Sect.~5.

\section{Observations and data reduction}
\label{data}
We used a set of images (Prop. 6864, P.I. Fusi Pecci) covering optical
to far-UV wavelengths, obtained with the Wide Field Planetary Camera~2
(WFPC2) on board the HST. The WFPC2 is a mosaic made of four
$800\times800$\, pixels cameras, with angular resolutions of
$0.046\arcsec/$pixel for the Planetary Camera (PC) and
$\sim0.1\arcsec/$pixel for the three Wide Field Cameras (WF2, WF3 and
WF4).  The optical dataset consists of $5 \times F555W$ images, three
with exposure time $t_{\rm exp}=100$\,sec and two with $t_{\rm
  exp}=7$\,sec, plus $4 \times F336W$ images with exposure time
$t_{\rm exp}=1600$\,sec.  The far-UV database comprises $3 \times
F160BW$ images with $t_{\rm exp}=1200$\,sec each. All the images have
the same pointing, with the cluster centre roughly centred in the PC
(see Figure~1). We combined the images using the IRAF task {\it
  imcombine} in order to improve the signal to noise ratio and
decontaminate them from cosmic rays, which are particularly prominent
in long exposure data.  For a given passband and exposure time we
adopted the resulting median frame as reference image for the data
reduction.  As done in our previous work (see for example Dalessandro
et al.~2009), the data reduction of the optical images was performed
using the ROMAFOT package (Buonanno et al.~1983), specifically
developed to perform accurate photometry in crowded regions (Buonanno
\& Iannicola~1984).  Because of under-sampling problems related to the
Point Spread Function (PSF) of stars in the far-UV images, we used
aperture photometry performed with the publicly available software
Source Extractor (SExtractor, Bertin \& Arnouts 1996).  The adopted
aperture radii were 2 pixels (corresponding to $0.2\arcsec$) for the
WF chips and 3 pixels ($\sim0.15\arcsec$) for the PC. For the WF
cameras, because of strong vignetting problems affecting the upper
right corners of all the $F160BW$ images, the regions $[550\lsim
  x\lsim800; 550\lsim y\lsim800]$ have been excluded from the
analysis. In each filter for every star detected, the photometric
error has been defined as the standard deviation of all the measures
obtained. 

The final star list includes all the sources detected in at least 2
filters.  Hot stars detected in $F160BW$ and not automatically found
in the optical data were force-fitted in the optical images. This
procedure added 10--15 HB stars in the F336W and F555W images. These
stars are typically a magnitude fainter than the TO level in the optical
CMD (Fig.~2).  This approach allowed us to sample in a proper way both
the cool sequences (like MS and RGB) thanks to the optical bands, and
the very hot ones (extremely blue HB stars) because of the sensitivity
of the far-UV bands to high temperatures.

Our photometric strategy assured a very high degree of
completeness in our samples.  In particular, selecting stars where
they are the brightest and then finding the corresponding stars in the
plane where they are faint, guarantees an high completeness level in
each plane.  The F160BW images are not affected at all by crowding
problems, allowing a quite trivial detection of all hotter objects.
As a consistency check we counted the intermediate-temperature HB
samples in both the optical and UV planes. The samples are virtually
identical (240 stars in the optical and 243 in the UV catalogue).

\begin{figure}
\includegraphics[width=84mm]{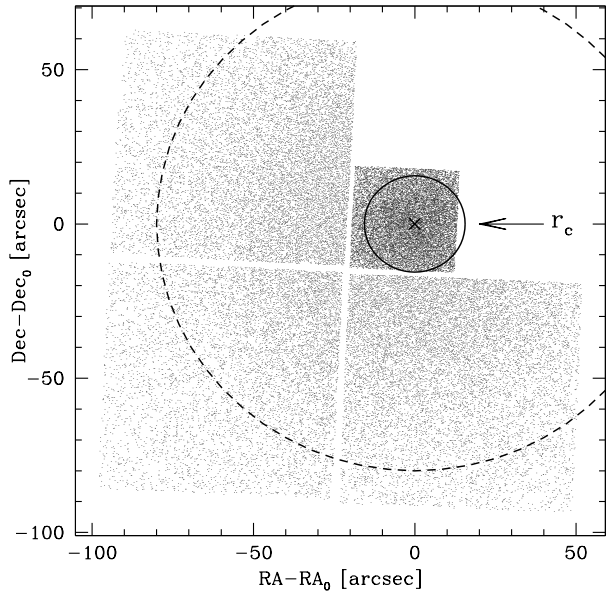}
\caption{Map of the WFPC2 data. The black cross corresponds to the cluster centre of gravity 
as determined in Sect.~\ref{astro}. The solid circle corresponds to the core radius
($r_c=15.6\arcsec$; Harris 1996), while the dashed circle
($r=75\arcsec$) defines the area within which we limited our analysis.}
\label{map}
\end{figure}
 
\begin{figure}
\includegraphics[width=84mm]{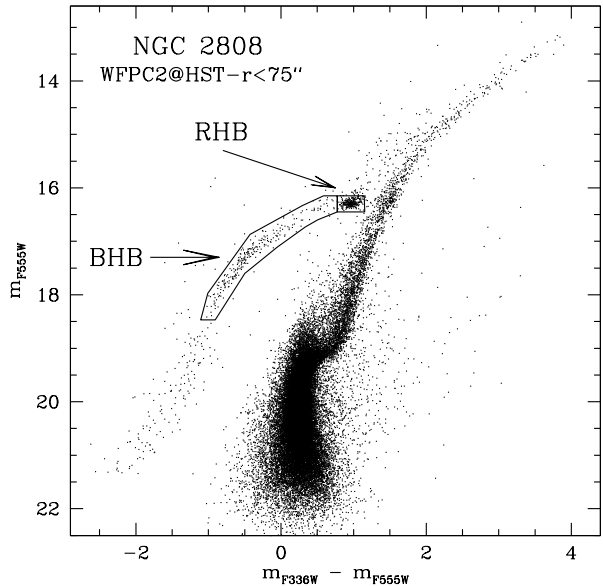}
\caption{Optical ($m_{F555W}, m_{F336W}-m_{F555W}$) CMD. We also show the boxes used to select RHB and BHB
stars.}
\label{cmd}
\end{figure}
  
\subsection{Photometric calibration, astrometry and centre of gravity}
\label{astro}
The WFPC2 photometric catalogue has been placed onto the absolute
GSCII astrometric system by using the procedures described in
Ferraro et al.~(2001, 2003). To maximize the number of stars in the
WFPC2 Field of View (FOV) adopted as reference for the
roto-traslation procedures, we used the photometric catalogue
described by Fabbri et al.~(2010) that was obtained with the Wide
Field Imager (WFI) at the 2.2 m ESO-MPI La Silla telescope, as
secondary astrometric standard list. With the help of the {\it
CataXcorr} package, developed by Paolo Montegriffo at the Bologna
Astronomical Observatory, we found some hundreds of stars in common
between the WFPC2 and the WFI catalogues. These were used to obtain a
very accurate astrometric solution. At the end of the procedure the
typical error is $\sim0.2\arcsec$ both in Right Ascension ($\alpha$)
and Declination ($\delta$).

All the magnitudes were transformed to the VEGAMAG photometric system
using the procedure described in detail by Holtzman et al.~(1995) with
the gain settings and ZeroPoints listed in Tab. 28.1 of the {\it HST
data handbook}.  The resulting CMDs are shown in Figs.~2 and ~3.

We determined the centre of gravity ($C_{\rm grav}$) of NGC~2808 with an
iterative procedure (Montegriffo et al.~1995) by
averaging the absolute positions ($\alpha$ and $\delta$) of stars
lying within $r=10\arcsec$ from a given point, starting from the centre reported by
Harris~(1996).  As done in previous papers (see Lanzoni et al~2007 for
an example) we selected four different sub-samples of stars with
different magnitude limits ($m_{F555W}=20$, 19.5, 19 and 18.5) to avoid incompleteness
effects in the central region. The average of the four derived values has been adopted as
$C_{\rm grav}$ and it
is located at $\alpha_{\rm J2000}=9^{{\rm h}}:12^{{\rm m}}:03.06^{{\rm
s}}$ and $\delta_{\rm J2000}=-64^{\circ} 51\arcmin 48.82\arcsec$,
 $\sim3.5\arcsec$ northeast from the
centre quoted by Harris (1996).

\begin{figure}
\includegraphics[width=84mm]{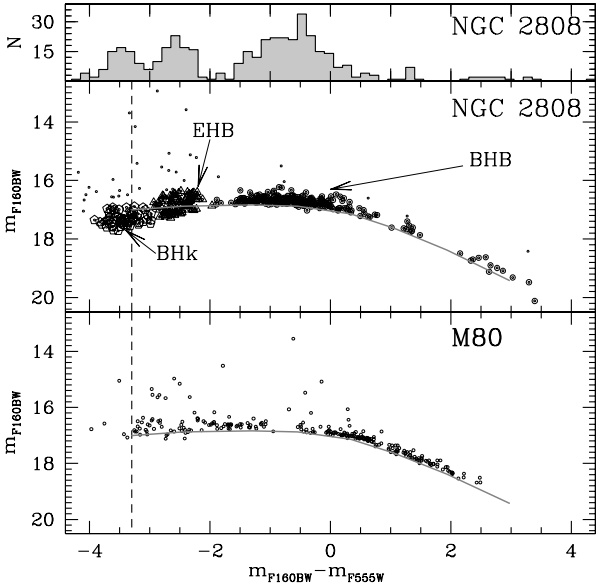}
\caption{{\it Bottom panel}. Far-UV ($m_{F160BW}$, $m_{F160BW} -
  m_{F555W}$) for M80 which we adopt as a ``normal'' blue tail
  cluster. Data are from Ferraro et al. (1998). The solid grey line is
  a fiducial line that reproduces the lower envelope of the HB star
  distribution. The colour and magnitudes have been shifted to
  compensate for the difference in reddening and distance modulus
  between M80 and NGC~2808 as determined by demanding that lower envelopes of the
  two BHBs match in the interval $-1<(m_{F160BW}-m_{F555W})<2$. {\it
  Middle-panel}. The far-UV CMD for NGC~2808. Different symbols denote
  different HB sub-populations: open circles are BHB stars, open
  triangles are EHB stars and open pentagons are candidate BHk stars.
  The dashed vertical line marks the hot edge of the distribution of
  HB stars in M80.  {\it Upper upper panel}. Far-UV star count
  distribution of NGC~2808.}
\label{cmdfuv}
\end{figure}

\section{The observed horizontal branch populations}
\label{obs}

In the canonical view, the HB is expected to be populated by stars
with different masses, decreasing from cold to hot effective
temperatures, all undergoing central He-burning and shell H-burning.
Given that the mass of the He-core at the start of the HB phase (Zero
Age Horizontal Branch--ZAHB) is almost constant
($M_{c}\sim0.5M_{\odot}$) and equal to the core mass at the He-flash,
the location of each star along the HB is essentially determined by
the mass of the surrounding envelope that survives the mass-loss
episodes during the RGB phase.  From the observational point of view,
the first physical parameter affecting the HB morphology is the
metallicity. Metal rich clusters usually have red HBs, while the
metal-poor ones tend to have HBs populated at bluer colours.  However,
at least one additional, or second, parameter is required to explain
the observed colour distributions of HB stars in Galactic GCs. This
second parameter problem has received much attention during the last
four decades (see for example Gratton et al. 2010 and references
therein).  Age has been the most commonly discussed second parameter
(Lee et al. 1994; Dotter et al. 2010; Gratton et al. 2010).  Very
recent work by Gratton et al. (2010) and Dotter et al. (2010) shows
that age as a global second parameter can account for general HB
parameters such as mean colour and length of the HB. However, HB
morphologies show a diversity which cannot be fully characterized by
mean properties. Sets of clusters with similar metallicities
(M3/M13/M80, see Ferraro et al.~1998; 47~Tuc/NGC~6388/NGC~6441, Rich
et al.~1997) have HBs distinguished by gaps and long blue
tails. Features like these have prompted discussion of additional
parameters like mass-loss efficiency during the RGB (Rood et al. 1993;
Ferraro et al. 1998; Dotter 2008) or initial He abundance differences
(D'Antona et al. 2002). The situation is still more complicated because
deviations from the general behaviour arise due to internal second or
third parameters that may vary from cluster to cluster. In this
context NGC2808 is a particulary interesting case which certainly
requires an accurate and detailed comparison between observations and
state-of-the-art theoretical models. As shown in Figs~\ref{cmd} and
\ref{cmdfuv}, the HB of NGC~2808 is characterized by a well populated
red clump and, on the blue side, a well populated and extended blue
sequence reaching down to $m_{F555W}\sim 21.5$, almost 2 magnitudes
fainter than the MS turn-off (TO).

By combining observations in optical and far-UV bands we can sample HB
stars according to their temperatures and spectral type. The strategy
adopted here is to select stars along the HB in the most appropriate
CMD according to their temperature. Hence, the red clump stars have
been selected in the optical CMD ($m_{F555W}$, $m_{F336W}-m_{F555W}$),
that is more sensitive to relatively cool stars, while the hottest HB
stars were selected in the far-UV CMD, where they are the most
luminous sources.  These selection criteria allow also a more accurate
distinction between pure HB and post-HB stars.  The
intermediate-temperature HB population has been selected in both CMDs,
and it was used a link between the cool and the hot part of the HB, as
well as to define normalization criteria and the relative
incompleteness of the two samples.  Two under-populated narrow regions
(or ``gaps'') are apparent in Figures~2 and 3.  These gaps were
previously identified by Bedin et al. (2000; hereafter B00) at $V=18.6$
and $V=20$ and the location in our CMD (at $m_{F555W}=18.55$ and
$m_{F555W}=20.05$) is well consistent with previous estimates.  The
two gaps split the HB in 4 subgroups (if we also consider the standard
subdivision at the cool and hot sides of the RR Lyrae instability
strip), that were named RHB, EBT1, EBT2 and EBT3 by B00 (see also
Castellani et al.~2006; Iannicola et al.~2009, hereafter I09).  For sake of homogeneity with
previous work of our group, we will use the nomenclature introduced by
Dalessandro et al.~(2008), and we will denote the four sub-populations
as red HB (RHB), blue HB (HB), extreme HB (EHB) and blue-hook (BHk), respectively.
The physical properties of
these 4 sub-groups will be discussed in the following section.

Because of well known vignetting problems affecting the upper-right
corner of all F160BW images (see Sect.~\ref{data}), we restrict our
analysis to stars with distances $r<75\arcsec$ from $C_{\rm grav}$.
The selected populations are shown with different symbols in Figs.~2
and 3.  The total HB sample in our catalogue is 616 stars: 256 RHB,
243 BHB, 66 EHB and 51 BHk sources.  They correspond, respectively, to
$(41\pm3)\%$, $(39\pm3)\%$, $(11\pm2)\%$ and $(9\pm1)\%$ of the total.
These ratios are in agreement with those reported by I09 and are
compatible with those quoted by B00. We
found 4 RR~Lyrae in common between the variable star list published by
Samus et al. (2009) and Corwin et al. (2004) and our BHB sample.
Given the small number, their inclusion doesn't effect the comparison
with theoretical models performed in the following sections, nor the
results obtained in this work.

To fully appreciate the oddity of NGC~2808's HB, it is useful to
compare it to a ``normal'' blue tail cluster, M80. Of the clusters
we've studied in the far-UV, M80 has the most extended BT. Its CMD
along with that of NGC~2808 is shown in Figure~\ref{cmdfuv}. To
facilitate comparison the colour and magnitude have been shifted to
compensate for the difference in reddening and distance modulus
between M80 and NGC~2808 as determined by demanding that lower
envelopes of the two BHBs match in the interval
$-1<(m_{F160BW}-m_{F555W})<2$ and added a fiducial. The fiducial marks
the lower envelope of the M80 HB.\footnote{Using ZAHBs of different
  metallicity we have determined that it is reasonable to use the
  shifted M80 fiducial for NGC~2808 despite the difference in
  metallicity.} There are two dramatic differences:
first, the HB of NGC~2808 is more extended toward high temperatures
than that of M80; second, if the fiducial is linearly extrapolated to
higher temperature many of the NGC~2808 stars lie below the
fiducial. To fit the observed NGC~2808 sequence the fiducial would have
to hook downward. While these differences become 
evident in far-UV CMDs, they are barely recognizable in purely optical 
diagrams.

\begin{figure}
\includegraphics[width=84mm]{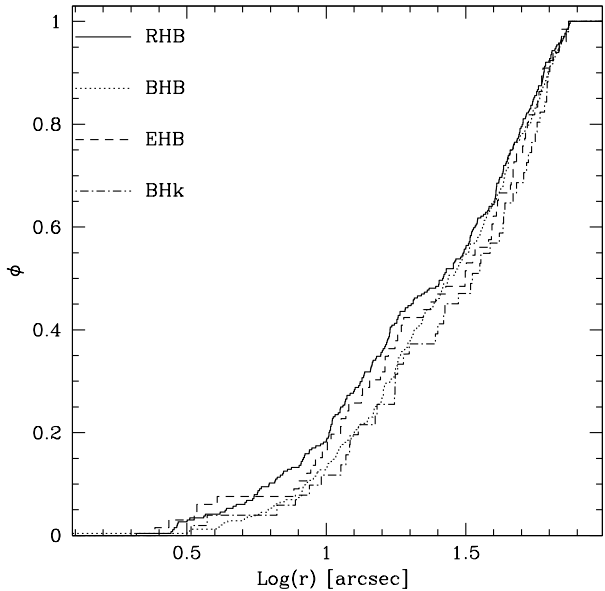}
\caption{Cumulative radial distribution of the four different HB sub-populations.}
\label{ks}
\end{figure}

The spatial distribution of a stellar population can give
information about its origin. Figure~\ref{ks} shows the cumulative
radial distributions of the 4 HB subgroups. The
RHB, BHB, EHB and BHk stars have similar radial distributions within the
WFPC2's FOV. A Kolmogorov-Smirnov (KS) test shows that there are no
statistically significant differences between the radial distributions
of these 4 sub-populations.  The strongest discrepancy is found between
RHB and BHB stars, but it is significant only at the $\sim1.3\sigma$
level.  I09 and B00 found similar results over the area covered by our
data. I09 did find significant changes in the ratios of the different
HB subgroups between the centre and their outer zone ($r>90\arcsec$).

\section{Theoretical analysis}
\label{theory}

We have compared our photometry with an appropriate set of BaSTI
$\alpha$-enhanced HB tracks and isochrones (Pietrinferni et al.~2006)
for ${\rm [Fe/H]}=-1.31$ close to the old Zinn \& West (1984) value of
$-1.36$. Current spectroscopic estimates are larger (${\rm
[Fe/H]}=-1.18$, Rutledge et al.~1997; ${\rm [Fe/H]}=-1.10$ or $-1.15$,
Carretta et al.~2009). A change of 0.2 in [Fe/H] shifts a ZAHB by 0.08
mag. A comparison with the observations can be made with a
compensating shift in distance modulus. 

We considered the entire set
of initial He abundances available in the BaSTI database,
i.e. Y=0.248, 0.300, 0.350 and 0.400.

\begin{figure}
\includegraphics[width=84mm]{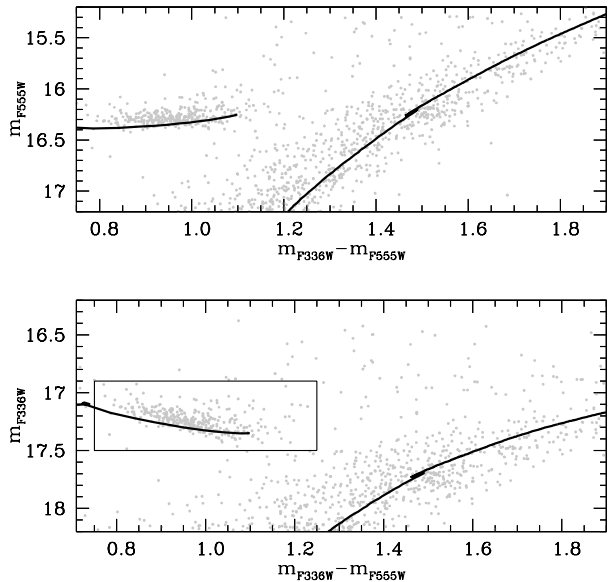}
\caption{($m_{F555W}$, $m_{F336W}-m_{F555W}$) and ($m_{F336W}$, $m_{F336W}-m_{F555W}$) 
CMDs for the cluster RHB population, and a portion of the cluster RGB. 
The box in the lower panel marks the region used in the comparison with synthetic HB diagrams (see text for details).  
The solid lines are the theoretical ZAHB and the isochrones at $t=12$Gyr 
and [Fe/H]$=-1.31$  with $Y=0.248$, 
shifted by the distance and extinction determined for the cluster.}
\label{teo1}
\end{figure}

The bolometric corrections to the WFPC2 bands have been determined
using the $\alpha$-enhanced spectral library by Castelli \&
Kurucz~(2004)---the same one adopted by Pietrinferni et
al.~(2006)---and following the procedure outlined by Girardi et
al.~(2002). This method allows a straightforward calculation of the effect of
extinction in the adopted filters, and its dependence on both
effective temperature ($T_{\rm eff}$) and surface gravity. We adopted
the Cardelli, Clayton \& Mathis~(1989) extinction law, with $R_V=3.1$.

The spectroscopic study by Pace et al.~(2006) showed that in NGC~2808
HB stars hotter than $\sim12000$\,K have surface [Fe/H], [Cr/H] and
[Ti/H] values around solar or even higher (up to ${\rm [X/H]}\sim1$,
where X is the abundance of one of these three metals) due to
radiative levitation in the atmosphere. In the ZAHB models and HB
tracks here adopted we mimic this effect by applying bolometric
corrections appropriate for ${\rm [Fe/H]}=0.0$ when $T_{\rm eff}$ is
between 12,000~K and 13,000~K, and, when the temperature is higher,
bolometric corrections appropriate for ${\rm [Fe/H]}=0.5$, based on
the spectroscopical evidence provided by Pace et al.~(2006, see their
Figure~4).  

This is in principle a crude approximation, but extended
(in both mass and initial chemical composition) grids
of HB stellar
evolution and atmosphere models that include consistently the
effect of radiative levitation
(see , e.g. Hui-Bon-Boa, LeBlanc \& Hauschildt~2000, and Michaud,
Richer \& Richard.~2008 for first
results on this topic) are still lacking. Also, it appears that
additional mixing processes (e.g. meridional circulation) must be
coupled to the effect of radiative levitation, in order to
explain the abrupt disappearance of the surface abundance anomalies
at $T_{eff} $11000-12000~K (see, e.g., Quievy et al.~2009).

It is clear that our way to simulate the effect of radiative levitation
is just an approximation, that cannot take into account a possible
stratification of the metal abundances in the atmosphere and 
below the photosphere.
In the following we will however show how this treatment 
seems to be reasonably appropriate.

\begin{figure}
\includegraphics[width=84mm]{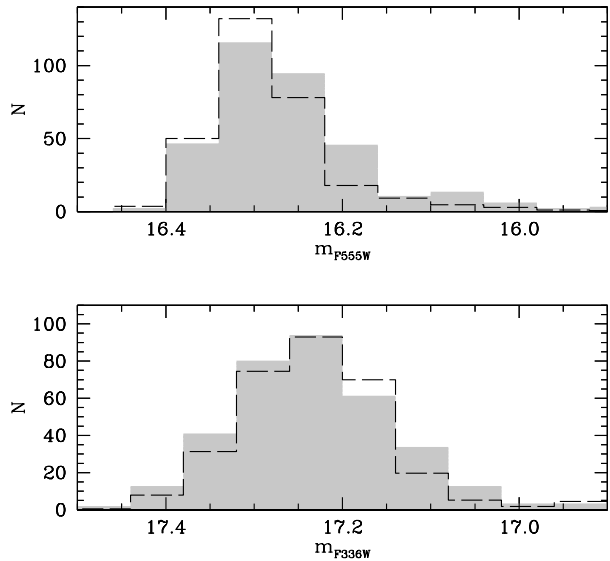}
\caption{Comparison between observed (grey histograms) and theoretical
  (dashed histograms) 
star counts for the RHB population,  
as a function of $m_{F336W}$ and $m_{F555W}$ magnitudes.}
\label{teo2}
\end{figure}

\subsection{Determination of the cluster reddening and distance modulus}
\label{distance}

As a first step we determined a reference distance modulus and
extinction by fitting our HB models to the RHB. First, we produced
synthetic samples of HB stars by following the techniques pioneered by
Rood~(1973; see also Ferraro et al.~1999) and compared separately the 
magnitude distributions of synthetic RHB
stars in the $F336W$ and $F555W$ filters
to the observed ones for the stars populating the box
displayed in the lower panel of Figure~\ref{teo1}. RHB stars are too
faint in the $F160BW$ filter to provide any additional constraint.

We used the set of models with $Y=0.248$ for the RHB stars.
The synthetic samples have four free parameters: extinction, distance
modulus $(m-M)_{0}$, mean HB stellar mass $\langle M_{\rm HB} \rangle$
and its dispersion $\sigma_{\langle M_{\rm HB} \rangle}$.  We started
by randomly selecting the stellar mass $M_{\rm HB}$ from a Gaussian
distribution centred around a given guess value of $\langle M_{\rm HB} \rangle$,
with 1$\sigma$ dispersion $\sigma_{\langle M_{\rm HB} \rangle}$.  The
WFPC2 magnitudes of the synthetic star were determined according to
its position along the appropriate HB track with mass $M_{\rm HB}$
(interpolated, when necessary, among the available set of tracks in
the BaSTI database) after an evolutionary time $t$. We determined $t$
assuming that the stars reach the ZAHB with a constant rate. We employed a
flat probability distribution ranging from zero to $t_{\rm HB}$, where
$t_{\rm HB}$ is the time spent along the HB (we consider as the end of the HB
phase the time when the central He-abundance drops
to zero).  The star with the lowest mass has the longest central
He-burning lifetime (Castellani et al.~1994, Cassisi et al.~2003) and
sets $t_{\rm HB}$.  This implies that for some masses the randomly
selected value of $t$ will be longer that their HB lifetime, or in other words
that they have already evolved to the following evolutionary stages.

\begin{figure}
\includegraphics[width=84mm]{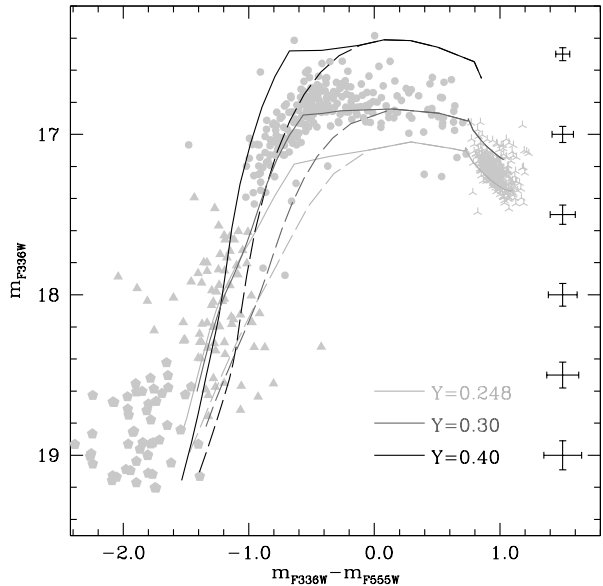}
\caption{Optical CMD of the cluster HB compared to theoretical ZAHBs 
for Y=0.248, Y=0.30 and Y=0.40 (see labels).
The dashed lines show the effect of neglecting radiative levitation 
in the ZAHB bolometric corrections when $T_{\rm eff}$ is larger 
than 12000~K (see text for details). The typical behaviour of the photometric errors
is also shown as a function of magnitude.}
\label{teo4}
\end{figure}

We modifed these synthetic magnitudes accounting for distance modulus
$(m-M)_0$ and extinction. We used as broad guidelines the results from
previous investigations (e.g. Bedin et al.~2000), exploring a range
of extintion values $E(B-V)=0.15$--0.20, and distance
modulus $(m-M)_0=15.0$--15.5. We finally added a Gaussian random error
with 1$\sigma$ dispersion, equal to that obtained from the data reduction
process. We then compared the resulting synthetic distribution
of $m_{F555W}$ and $m_{F336W}$ values to the observed one, 
and modified the values of the four free parameters until 
the mean magnitude and 1$\sigma$ dispersion equal the observed 
ones in both $m_{F555W}$ and $m_{F336W}$.

For each choice of the four free parameters we have produced 200 synthetic
distributions, each of them with the same number of objects as in the
observational sample. The synthetic star counts as a function of magnitude
displayed in the rest of the paper are average values from these ensembles
of simulations.

We found $A_{F555W}=0.52$ (corresponding to $E(B-V)=0.17$), 
$(m-M)_0=15.23$, $\langle M_{\rm HB} \rangle=0.69 M_{\odot}$ and 
$\sigma_{\langle M_{\rm HB} \rangle}=0.015 M_{\odot}$.

The theoretical ZAHB and the RGB of a 12~Gyr isochrone with $Y=0.248$ 
shifted according to the best-fit distance modulus and
extinction are shown in Figure~\ref{teo1}. 
Figure~\ref{teo2} shows the magnitude distributions for the RHB stars in the 
observed (shaded histogram) and synthetic (dashed histogram), for the best fit parameters
obtained above.  We also checked---comparing theoretical and
observed star counts along the RGB---that the magnitude of the
so-called RGB bump is well reproduced in both $m_{F555W}$ and
$m_{F336W}$ by the 12~Gyr theoretical isochrone, for the combination
of distance modulus and extinction determined from the HB.

By considering the initial mass of the objects populating the tip of
the RGB in the 12~Gyr isochrone, we found that the mean amount of mass
lost along the RGB by the $Y=0.248$ population is $\langle \Delta M
\rangle=0.15\,M_{\odot}$. This value is very similar to the results of the synthetic
modeling of the RHB stars by D'Antona \& Caloi~(2004), who obtained
$\langle \Delta M \rangle=0.13\,M_{\odot}$ and a dispersion
$\sigma_{\langle M_{\rm HB} \rangle}=0.015\,M_{\odot}$, using their
scaled solar isochrones with approximately the same [Fe/H] of our
$\alpha$-enhanced models (and slightly lower $Y$).

\begin{figure}
\includegraphics[width=84mm]{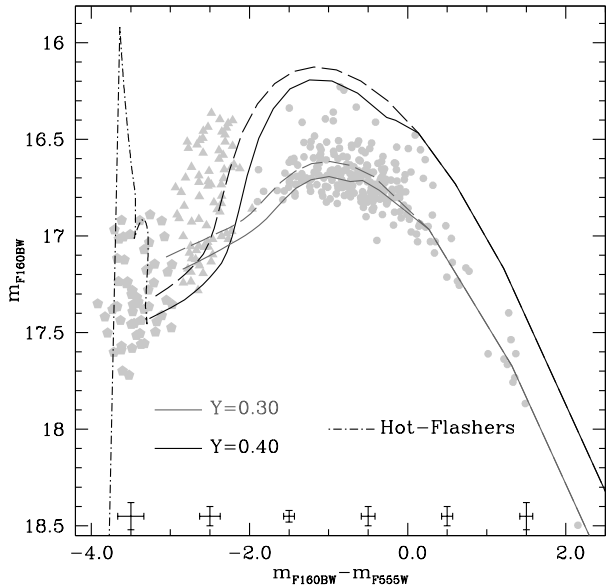}
\caption{As in Fig.~\ref{teo4}, but for the ($m_{F160BW}$, $m_{F160BW}-m_{F555W}$) CMD. The RHB population is not displayed 
because too faint in the F160BW filter. 
The dashed-dotted line represents the hot-flasher model used to reproduce the BHk stars, as described in
Section~\ref{bhk}. The behaviour of the photometric errors as a function of colour is also shown.}
\label{teo5}
\end{figure}

\subsection{Multiple-He populations along the HB}

To investigate the nature of the BHB, EHB and BHk populations we
used the most appropriate CMDs: the ($m_{F336W}$, $m_{F336W}-m_{F555W}$) CMD 
has been used as a bridge between the
RHB and BHB populations, and the ($m_{F160BW}$, $m_{F160BW}-m_{F555W}$)
CMD to characterize the BHB, EHB and BHk populations. Prompted by the
multiple MS detected in the cluster (Piotto et al.~2007) we
used models with different He content for various components of the
HB.  Figure~\ref{teo4} shows the
($m_{F336W}$, $m_{F336W}-m_{F555W}$) CMD of RHB and BHB stars. 
Using the distance modulus and extinction obtained in the previous
section, we also display three ZAHBs with different Helium
abundances $Y=0.248$, 0.30 and 0.40 respectively. For each ZAHB we show
two sequences, one that mimics the inclusion of levitation as
described above (solid line), and one without levitation (dashed
line).  We see that:

\begin{enumerate}
\item{Applying metal-rich bolometric corrections to mimic 
radiative levitation significantly affects the
location of the ZAHB in the high temperature regime, i.e. for
$(m_{F336W}-m_{F555W})<0$. In particular, the ``levitation'' ZAHB
is systematically bluer ($\delta (m_{F336W}-m_{F555W})\sim 0.4$)
than the ``no-levitation'' ZAHB (or the ZAHB level is brighter
at fixed colours).}  

\item{The BHB stars define a sort of ``knee'' in the stellar
distribution and are significantly brighter than the RHB clump in this
plane. Figure~\ref{teo4} shows very clearly that even including
radiative levitation, the magnitude level of the BHB sub-population is
not reproduced by the ZAHB with canonical helium abundance
($Y=0.248$): the BHB appears significantly (0.3--0.4 mag)
brighter than the canonical ZAHB.  On the other hand, the ``levitation'' ZAHB
with $Y=0.30$ matches very nicely the lower envelope of the BHB star
distribution.}

\item{The cooler BHB stars, $(m_{F336W}-m_{F555W})>0$, are not affected
  by levitation, yet they still lie significantly above the canonical
  $Y=0.245$ ZAHB. They are well fit by the $Y=0.30$ ZAHB. The
  conclusion that $Y$ for the BHB is $\sim 0.05$ greater than $Y$ of the
  RHB is independent of levitation and our treatment thereof.}

\end{enumerate}

The situation is less clear for EHB and BHk stars because their
distribution becomes almost vertical in this CMD, and photometric
errors increase significantly the scatter at faint magnitudes.  For
these reasons we used the ($m_{F160BW}$, $m_{F160BW}-m_{F555W}$)
CMD in order to study the properties of these sub-populations.
Figure~\ref{teo5} shows the CMD of the three bluest sub-populations,
together with the ZAHB for different initial values of $Y$, with and
without inclusion of levitation.  It is interesting to note that
in the F160BW filter the inclusion of levitation goes in the opposite
direction than in optical filters, making the ZAHB fainter
at a given colour. Additional interesting features emerge from this
figure:

\begin{enumerate}
\item{Just as it was in the ($m_{F336W}$, $m_{F336W}-m_{F555W}$) CMD,
the BHB population is matched by a ZAHB with $Y=0.30$ including the
effect of levitation.}  
\item{The EHB population is consistent with a ZAHB with $Y=0.40$
including levitation.}  
\item{The BHk stars are too hot to be reproduced by any ZAHB models,
hence they require a more separate discussion (see Section 4.3)}
\end{enumerate} 

From this comparison we can safely conclude that three (out of four)
of the sub-populations along the HB of NGC~2808 can be identified as
the progeny of three distinct stellar populations with different
initial helium abundances in the range $Y=0.24$--0.40. The specific
values used ($Y=0.248$, 0.30, 0.40) were set by the model data base,
and we have not attempted to determine values of $Y$ which best fit
the data. The adopted values are consistent with what found from the
photometry of the MS (Piotto et al. 2007).  This conclusion is
independent of any assumption about the mass-loss efficiency of their
RGB progenitors.  Our findings are also in agreement with He
abundances reported on the basis of different indicators along the RGB
(like difference in colour or $T_{\rm eff}$) by Bragaglia et
al. (2010).  The consistency with these results also justifies our
choice (in principle arbitrary) to fix the distance (and extinction)
by matching the $Y=0.248$ HB population to the RHB.  Had we tried to
match these ``normal-He'' models to, e.g. the BHB population, it would
have been impossible then to match the RHB stars with models with any
reasonable value of $Y$.

\begin{figure}
\includegraphics[width=84mm]{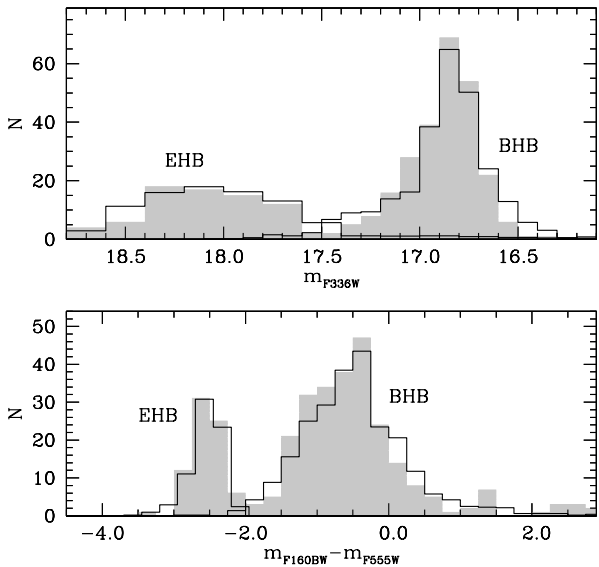}
\caption{Comparison between observed (grey histograms) and theoretical
  (solid histograms) star counts for the 
BHB (matched with models with $Y$=0.30) 
and EHB populations (matched with models with $Y$=0.40) 
as a function of the $m_{F336W}$ magnitudes (upper panel) and
($m_{F160BW}-m_{F555W}$) colour (lower panel).}
\label{teo6}
\end{figure}

As a second step of our analysis, we determined the mean mass of
the BHB and EHB sub-populations, by producing synthetic HBs in the same
way as described before, and matching the mode of both
the $m_{F336W}$ and the ($m_{F160BW}-m_{F555W}$) distributions.  The synthetic sample
compared to the BHB component has been calculated using the $Y=0.30$
HB tracks, while for the EHB we employed the $Y=0.40$ tracks,
including the effect of levitation in both cases.  Figure~\ref{teo6}
compares the observed (shaded histogram) and theoretical (solid histogram)
star counts as a function of $m_{F336W}$ and the
($m_{F160BW}-m_{F555W}$) for the BHB and the bluer EHB component.
An indirect estimate of mass-loss experienced by each sub-population
can then be obtained by comparing the mean mass on the HB and the mass
at the TO (for an age of 12\,Gyr). A summary of the values obtained for
RHB, BHB and EHB sub-populations is reported in Table~1.
Under these assumptions we found a mean amount of RGB mass-loss ranging from 
$\langle \Delta M \rangle=0.15$  for both the RHB and EHB stars,
to $\langle \Delta M \rangle=0.20$ for the EHB stars. 

Finally, in order to verify whether the RHB, BHB and EHB populations
are the progeny of the red, mean and blue MS, respectively, we have
compared the star counts along the HB, with those reported by Piotto
et al.~(2007) along the MS.  In our sample we found: $(N_{\rm
RHB}/N_{\rm BHB})_{\rm obs}=1.1 \pm 0.1$, and $(N_{\rm RHB}/N_{\rm
EHB})_{\rm obs}=3.9 \pm 0.7$. Starting from the observed MS number
counts, and after applying corrections accounting for the different
evolutionary lifetimes along the RGB and among the HB populations 
(i.e., how the lifetime varies as a function of $Y$ and mass) the expected number of
stars along the HB would be: $(N_{\rm RHB}/N_{\rm BHB})_{\rm exp}= 3.7
\pm 0.4$ and $(N_{\rm RHB}/N_{\rm EHB})_{\rm exp}=3.3 \pm 0.5$.
Clearly, while the two values of $(N_{\rm RHB}/N_{\rm EHB})$ are
compatible, it is not possible to reconcile the observed and the
expected values of $(N_{\rm RHB}/N_{\rm BHB})$.  The ACS data used by
Piotto et al. (2007) are strongly off-centred ($r \sim 200\arcsec$)
with respect to $C_{\rm grav}$. As shown by I09, the ratio $N_{\rm
RHB}/N_{\rm BHB}$ roughly doubles for $r>90\arcsec$.  This may be a
possible explanation for the observed discrepancy. However the
progenitor masses for RHB and BHB are respectively $0.84\,M_\odot$ and
$0.76\,M_\odot$ (see Table~1). Dynamical evolution of the cluster
should cause the $N_{\rm RHB}/N_{\rm BHB}$ to decrease as $r$
increases, just the opposite of what was found by I09. A more detailed
study of radial variations in HB sub-populations extending over the
full cluster could be illuminating. It would also be quite interesting
to check whether the triple MS varies with position in the cluster.

\begin{figure}
\includegraphics[width=84mm]{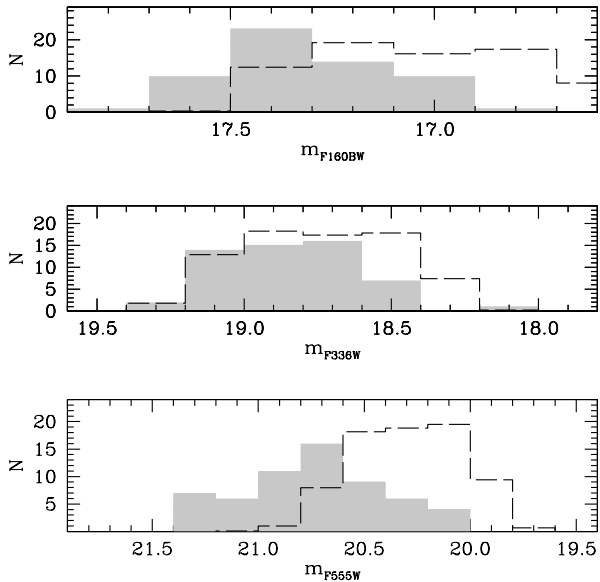}
\caption{Comparison between observed (grey histograms) and theoretical
  (dashed histograms - models 
with $Y$=0.40) star counts for the 
BHk population. The synthetic sample reaches the bluest possible
colours allowed by 
the theoretical models.}
\label{teo7} 
\end{figure}

\subsection{The Blue Hook population}
\label{bhk}

The CMDs in Figs.~\ref{teo4} and \ref{teo5} show very clearly that
BHk stars are not reproduced by any of the theoretical models, this
sub-population is systematically hotter than the hottest point along
the ZAHBs. 
To shed more light on the nature of these stars, 
in Fig.~\ref{teo7} we compare the observed number count
distributions for the BHk sub-population in the $F160BW$, $F336W$ and
$F555W$ magnitudes, with a synthetic sample with $Y=0.40$, a
mean mass $\langle M_{\rm HB} \rangle=0.461 M_{\odot}$ (i.e. the
minimum possible mass of HB objects for our models at this
metallicity) 
and zero dispersion. It is impossible 
to match the observed distributions in
all three filters.

This inability of HB models to reproduce the BHk population suggests
that these are not genuine HB stars, and it leads us to identify these
objects as hot-flashers (D'Cruz et al. 1996; Brown et al.~2001, 2010; Cassisi et al.~2003;
Moehler et al.~2004).

Hot-flashers are stars that experience strong mass-loss during the RGB
phase, leave the branch before the occurrence of the He-flash, and
move quickly to the He-core white dwarf cooling curve, where they
experience a He-flash under conditions of strong electron degeneracy
in their core.  This scenario was envisaged earlier by Castellani \&
Castellani (1993) who named these peculiar stellar objects ``red giant
stragglers.''  Stars that ignite helium on the white dwarf cooling
sequence have a much less efficient H-burning shell, and will undergo
extensive mixing between the He-core and the H-rich envelope (Cassisi
et al.~2003).  This makes the progeny of hot-flashers hotter and
fainter than genuine HB stars (i.e. stars that ignite He along the
RGB), describing a kind of hook in UV CMDs that gives the name to this
class of objects.  The presence of such peculiar sources in NGC~2808
has been already noted by pure UV analysis by Brown et al.~(2001) and
Dieball et al.~(2005).  As shown in Figure~\ref{teo5}, the location of
BHk stars is indeed reproduced by a hot-flasher model (dashed-dotted
line) from Cassisi et al.~(2003) for a mass $M=0.489
M_{\odot}$.\footnote{The bolometric corrections used
in the hot-flasher model displayed in Figure~\ref{teo5} have been
obtained from scaled-solar model atmospheres, accounting for the
effect of levitation as described in Section.~4. As shown by Brown et
al.~(2001), scaled solar spectra do not accurately represent the
peculiar atmospheres of hot-flashers, which are expected to have
strongly enhanced He and C abundances.  Therefore, for the hot-flasher
model one should in principle use more appropriate---but not yet
publicly available---bolometric corrections, that however, should not
alter the basic qualitative conclusions of our analysis.}

The use of hot-flasher models with different total mass would not
greatly modify the outlined scenario since all hot-flasher models
tend to cluster in a very narrow region of the CMD, regardless of
their total mass  (Brown et al.~2001; see also Miller Bertolami et al.~2008).
Also, the effect of the initial chemical
composition on the location in $T_{\rm eff}$ is minor, at least at
subsolar metallicities (see Miller Bertolami et al.~2008).
As a consequence it is not trivial to understand which $Y$
sub-population produces the BHks.
Potentially star counts derived from MS analysis might shed light on this point.
Hypothesizing three different progenitors for BHks we obtain
$(N_{\rm RHB+BHk}/N_{\rm BHB})_{\rm obs}=1.3 \pm 0.1$,
$(N_{\rm RHB}/N_{\rm BHB+BHk})_{\rm obs}=0.9 \pm 0.1$ and $(N_{\rm
  RHB}/N_{\rm EHB+BHk})_{\rm obs}=2.2 \pm 0.4$. In all cases adding
  the BHk stars to a given HB sub-populations makes the disagreement
  with the ratios expected from Piotto et al. (2007) worse.

 Bailyn (1995) has suggested that hot HB stars in a number of
 clusters have a binary origin. On the other hand a recent
 spectroscopic analysis (Moni Bidin et al.~2006) of a sample of hot HB
 stars in NGC~6752 found no close binaries. Neither of these studies
 have direct application to the case of NGC~2808. Bailyn's arguments
 apply to sdB stars which are probably EHB stars (Dorman et
 al. 1993). Our preliminary results indicate that NGC~6752 has no BHk
 stars, so the Moni Bidin result also applies to EHB stars. Our
 results here clearly show that EHB and BHk stars are not the same
 thing. One must be quite precise in defining these hot
 populations. Specifically in NGC~2808, Piotto et al.~2007 tentatively
 suggest that their observed binary sequence might be connected with
 the BHk stars, and they note that the binary fraction and BHk fraction
 are similar. Most GCs have binary stars, and most of these contain
 exotic populations like BSS which are produced by binary
 evolution. On the other hand, BHk stars are found in only a very few
 GCs. If binaries are producing BHk stars, they should be found in
 most GCs. So far genuine BHk stars have been found only in the most massive
 clusters most of which have multiple populations. This must be telling
 us something about their origin.

\begin{figure}
\includegraphics[width=84mm]{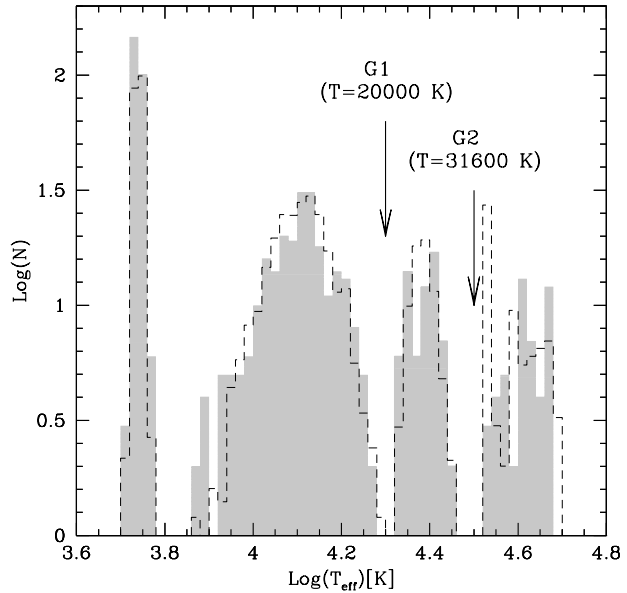}
\caption{The observed distribution of $T_{\rm eff}$ for NGC~2808 HB stars. Grey histograms are 
the temperature distribution for observed stars obtained as discussed in
Section~\ref{temperature}. Dashed black lines rapresent the temperature distribution of
simulated stars. The four well separated groups of stars are matched
with models with, respectively -- moving from the lowest to the highest
$T_{eff}$ -- $Y$=0.248, 0.30, 0.40, and the adopted track of hot-flashers.}
\label{isto}
\end{figure}

\subsection{The HB temperature distribution}
\label{temperature}

The combined use of far-UV and optical filters allows us to properly
visualize the different HB sub-populations and reveal the existence of
gaps, avoiding artificial features generated by the
saturation of the chosen colour index as it becomes insensitive to
changes in $T_{\rm eff}$.  The bottom panel of Figure~\ref{teo6} shows,
for example, how the HB distribution in the ($m_{F160BW}-m_{F555W}$) colour
separates very clearly the EHB and BHB sub-populations. 
In order to obtain the distribution in temperature of the different HB groups,
we first derived $T_{\rm eff}$ from the observed and dereddened  colours
by interpolating a spline along the colour-$T_{eff}$ relation given by the appropriate
theoretical ZAHBs.
The result is shown as grey histogram in Figure~\ref{isto}.\\
This approach neglects the post-ZAHB evolutionary effects on the colour-$T_{eff}$ relation
which however are small.
In fact the derived distributions have been found to be fully compatible with those directly
obtained from the synthetic models (dashed histograms in Figure~\ref{isto}).\\
The $T_{\rm eff}$ distribution displayed in Figure~\ref{isto} appears
clearly multimodal, as expected.  Each peak observed in this
distribution belongs unequivocally to only one of the 4
sub-populations.  The effective temperatures spanned by the HB stars
in NGC~2808 cover the range $5,000<T_{\rm eff}<50,000$\,K.  The RHB
has a distribution peaked at $T_{\rm RHB}\sim5,400$\,K, the BHB peaks
at $T_{\rm BHB}\sim12,600$\,K, the EHB at $T_{\rm EHB}\sim24,000$K,
while BHk stars are centreed around $T_{\rm BHk}\sim40,000 $\,K. The
gaps G1 and G2, that we have found at $m_{F555W}=18.55$ and
$m_{F555W}=20.05$, are located at $T_{\rm G1}\sim20,000$\,K and
$T_{\rm G2}\sim31600$\,K, respectively.  The temperatures of these two
gaps ($T_{\rm G2}$ in particular) are slightly hotter than those
($\sim17,000 $\,K and $\sim25,000 $\,K) estimated by Sosin et
al.~(1997).  These latter values are based on models with a constant
and standard ($Y=0.236$) He abundance. Moreover, since they were
unable to fit both the optical and near UV data with the same distance
modulus and reddening values, they applied colour shifts (for example
$\Delta(B-V)=0.07$ towards the red in the optical CMD) that may
explain the discrepancies in the gaps' temperatures. Our estimate for
$T_{\rm G1}$ and $T_{\rm G2}$ are much more compatible with the
temperatures estimated by the spectroscopic observations of a sample
of EHB and BHk stars by Moehler et al.~(2004; see their Table~4 and
their Figure~2). Their BHk candidates have typically $T_{\rm
eff}>35,000$\,K, while the few EHB stars observed are distributed in
the temperature interval $20,000<T_{\rm eff}<30,000 K$.  Our estimated
highest temperature along the HB is substantially higher than the
value ($\sim 37,000 K$) reported by Recio-Blanco et al.~(2006) from optical CMDs.
This
once more demonstrates the difficulties of deriving the properties of
extremely hot HB populations from pure optical CMDs.  In the optical
CMD a star moves down the BT as the bolometric correction increases
with $T_{\rm eff}$ or the stellar luminosity decreases. For BHk stars
both are changing and optical colours cannot give a reliable $T_{\rm
eff}$.  However it is important to note also that the temperatures
estimated for BHks are affected by large uncertainties that are not
easily quantifiable.  They are due to the lack of a proper treatment
of the peculiar chemical composition of their atmospheres and to the
photometric errors that become larger in $m_{F555W}$.

\begin{table}
\caption{Mass (in solar units) at the MS-TO ($M_{TO}$) and mean value along the HB
($\langle M_{\rm HB} \rangle$), for each of the three HB sub-populations. 
$\Delta M$ represents an estimate of the
mean mass lost during the RGB phase. $(T_{eff})_{peak}$ is the effective temperature
of the peak of the distribution of each HB sub-population.}
\begin{tabular}{@{}|p{1.2cm}|*{6}{c|}|}
\hline \hline
    POP  &   $Y$  &   $\langle M_{\rm HB} \rangle$  & $ M_{\rm TO}$ &
    $\Delta M $  &($T_{\rm eff})_{\rm peak}$\\
\hline
    RHB      &   0.248  &   0.84   &  0.69  &	 0.15  &  5,400 \\
    BHB      &   0.30   &   0.76   &  0.565 &	 0.195 &  12,600\\
    EHB      &   0.40   &   0.627  &  0.479 &	 0.148 &  24,000\\
\hline    

\label{tab:masses}
\end{tabular}
\end{table}

\section{Summary}

We have analyzed the complex, multimodal HB of NGC~2808 as a prototype
for a series of studies of hot HB stellar populations.  We found that
a combination of optical and far-UV magnitudes is the most powerful tool
to efficiently analyze multimodal HBs with extremely blue components,
in terms of both comparisons with theoretical models and direct star
count analyses. By following this approach, the presence of He-enhanced 
populations along the blue
part of the HB can be inferred without making assumptions
about the RGB mass-loss. This cannot be done using only optical CMDs
because HB sequences with different initial $Y$  overlap for
$T_{\rm eff}>10,000$\,K, where the HB becomes almost vertical because
of the large increase of the bolometric corrections with $T_{\rm eff}$.
We have used here the ($m_{F336W}$, $m_{F336W}-m_{F555W}$) CMD to
study the RHB and BHB stars and the ($m_{F160BW}$,
$m_{F160BW}-m_{F555W}$) plane for hotter sub-populations (EHBs and
BHks). BHB stars have been used as ``reference'' population to check 
the completeness level of the
samples observed in the two different planes.

From the comparison of theoretical ZAHB sequences with observations,
we confirm that the peculiar HB of NGC~2808 cannot be reproduced by a
population with a single initial He abundance, instead we need the
combination of three different populations with different helium
abundances ranging from $Y=0.248$ to 0.40. Our models use $Y=0.248$,
0.30, 0.40 which are consistent with those inferred by Piotto et
al.~(2007) from the observed split of the cluster MS.  Using synthetic
HB simulations we find that the CMD of the RHB sub-population is well
matched by models with standard initial $Y$ values ($Y=0.248$ in this
case) and an average mass $\langle M_{\rm HB} \rangle =0.69
M_{\odot}$. The BHB stars can be reproduced by models with $Y=0.30$
and $\langle M_{\rm HB} \rangle=0.57 M_{\odot}$.  The EHB
sub-population is matched by models with with $Y=0.40$, that might be
the progeny of the blue-MS population.  These latter stars arrive on
the ZAHB with a very tiny H envelope, that results in surface
temperatures $T_{\rm eff}>20,000 $\,K, with a total mass $\langle
M_{\rm HB} \rangle=0.48 M_{\odot}$.  By comparing the average mass of
the three sub-populations along the HB with those at the RGB tip for a
12~Gyr isochrones, ${\rm [Fe/H]} = -1.31$ and the appropriate values
of $Y$, we find that the populations have experienced an average
mass-loss in the range $\langle \Delta M \rangle \sim 0.17 \pm
0.02~M_{\odot}$.  On the basis of HB evolutionary tracks and synthetic
HB modeling we find that the HB of NGC~2808 is one of the most
extended in temperature, with stars spanning the interval
$5,000<T_{\rm eff}<50,000$\,K. These temperatures, estimated from
stellar evolutionary models, are consistent with those derived from
spectroscopic observations by Moehler et al.~(2004), but significantly
larger than those determined from optical CMDs. 

UV magnitudes for stars with $T_{\rm eff} > 11,500$\,K are affected by
the levitation of heavy elements greatly increasing atmospheric
opacity. We have crudely modeled the effects of levitation by
employing bolometric corrections appropriate for metallicities much
higher than that of NGC~2808. The resulting models agree quite well
with the observations especially for the BHB. Still, because of this
approximate treatment our values for $Y$ and the masses of the HB
sub-populations are uncertain. However, our result that there are 3
populations with helium abundances spanning a range of roughly 0.15 in
$Y$ is quite robust. 

Even models with extremely large $Y$ abundances are not able to
reproduce the location of the BHk stars. This sub-population has been
interpreted here as hot-flashers, the result of He-flash ignition
along the He-core white dwarf cooling sequence. Why these stars are
present in a few massive clusters and absent in other clusters with
comparably long blue tails is a challenge to stellar evolution theory.

A comparison between HB and MS sub-populations relative ratios
reveals that while $(N_{\rm RHB}/N_{\rm EHB})_{\rm obs}$ is
compatible with $(N_{\rm RHB}/N_{\rm EHB})_{\rm exp}$, it is
non-trivial to explain the discrepancy between the values obtained for
$(N_{\rm RHB}/N_{\rm BHB})_{\rm obs}$ and $(N_{\rm RHB}/N_{\rm
BHB})_{\rm exp}$. The four sub-populations (RHB, BHB, EHB and BHk)
show the same radial trend for $r < 75\arcsec$, in agreement with
previous findings (see B00 and I09). However, there is some evidence
that the sub-population ratios might vary over larger radial distances
(I09). Studies of radial variation of the various populations covering
the entire cluster coupled with detailed modeling of dynamics would be
valuable.

This paper is the first in a series which will make a detailed
comparison between theoretical models and observations of blue tail HB
clusters. \\

We thank the referee Aaron Dotter for the useful comments that
improved the presentation of this work.
This research was supported by the Agenzia Spaziale Italiana (under contract ASI-
INAF I/016/07/0), by the Istituto Nazionale di Astrofisica (INAF, under contract PRIN-
INAF2008) and by Ministero dell'Istruzione, dell'Universit\'a e della Ricerca.
RTR is partially supported by STScI grant GO-11975. SC acknowledges the financial support of 
the Ministero della Ricerca Scientifica e dell'Universita' PRIN MIUR 2007:
\lq{Multiple stellar populations in globular clusters}\rq\, 
and the Italian Theoretical Virtual Observatory Project.

\label{lastpage}

\begin{thebibliography}{}

\bibitem[Bailyn(1995)]{1995ARA&A..33..133B} Bailyn, C.~D.\ 1995, ARA\&A, 33, 133 
\bibitem[(Bedin et al.(2000)]{}%
    Bedin, L. R., Piotto, G., Zoccali, M., Stetson, P. B., Saviane, I.,
    Cassisi, S., \& Bono, G. 2000, A\&A, 363, 159
\bibitem[Bedin et al.(2004)]{2004ApJ...605L.125B} 
    Bedin, L.~R., Piotto, G., 
    Anderson, J., Cassisi, S., King, I.~R., Momany, Y., 
    \& Carraro, G.\ 2004, ApJl, 605, L125 
\bibitem[Bertin \& Arnouts(1996)]{be96} 
     Bertin, E., \& Arnouts, S.\ 1996, A\&AS, 117, 393 

   
\bibitem[Bekki 
\& Freeman(2003)]{2003MNRAS.346L..11B} Bekki, K., \& Freeman, K.~C.\ 2003, MNRAS, 346, L11 


\bibitem[Bragaglia et al.(2010)]{2010arXiv1005.2659B} Bragaglia, A., 
Carretta, E., Gratton, R., D'Orazi, V., Cassisi, S., 
\& Lucatello, S.\ 2010, arXiv:1005.2659 


\bibitem[Byun 
\& Lee(1991)]{1991ASPC...13..243B} Byun, Y.-I., \& Lee, Y.-W.\ 1991, The Formation and Evolution of Star Clusters, 13, 243 

\bibitem[Brown et al.(1991)]{1991ApJ...376..115B} Brown, J.~H., Burkert, 
A., \& Truran, J.~W.\ 1991, ApJ, 376, 115 

\bibitem[Brown et al.(1995)]{1995ApJ...440..666B} Brown, J.~H., Burkert, 
A., \& Truran, J.~W.\ 1995, ApJ, 440, 666 

\bibitem[Brown et al.(2001)]{2001ApJ...562..368B} Brown, T.~M., Sweigart, 
A.~V., Lanz, T., Landsman, W.~B., \& Hubeny, I.\ 2001, ApJ, 562, 368 

\bibitem[Brown et al.(2010)]{2010arXiv1006.1591B} Brown, T.~M., Sweigart, 
A.~V., Lanz, T., Smith, E., Landsman, W.~B., 
\& Hubeny, I.\ 2010, arXiv:1006.1591 

\bibitem[Buonanno et al. (1983)]{buon83} Buonanno, R., Buscema, G., Corsi,
C.~E., Ferraro, I., \& Iannicola, G.\ 1983, A\&A, 126, 278

\bibitem[Buonanno \& Iannicola (1989)]{buon89}Buonanno, R., \& Iannicola,
G. \ 1989, PASP, 101, 294

\bibitem[Busso et al.(2007)]{2007A&A...474..105B} Busso, G., et al.\ 2007, A\&A, 474, 105


\bibitem[Calamida et al.(2009)]{2009ApJ...706.1277C} Calamida, A., et al.\ 
2009, ApJ, 706, 1277 

\bibitem[Cardelli et al.(1989)]{1989ApJ...345..245C} Cardelli, J.~A., 
Clayton, G.~C., \& Mathis, J.~S.\ 1989, ApJ, 345, 245 

\bibitem[Carretta et 
al.(2006)]{2006A&A...450..523C} Carretta, E., Bragaglia, A., Gratton, R.~G., Leone, F., Recio-Blanco, A., \& Lucatello, S.\ 2006, A\&A, 450, 523 

\bibitem[Carretta et al.(2008)]{2008arXiv0811.3591C} Carretta, E., 
Bragaglia, A., Gratton, R.~G., \& Lucatello, S.\ 2008, arXiv:0811.3591

\bibitem[Carretta et 
al.(2009)]{2009A&A...508..695C} Carretta, E., Bragaglia, A., Gratton, R., D'Orazi, V., \& Lucatello,
S.\ 2009, A\&A, 508, 695 

\bibitem[Casetti-Dinescu et al.(2007)]{2007AJ....134..195C} 
Casetti-Dinescu, D.~I., Girard, T.~M., Herrera, D., van Altena, W.~F., 
L{\'o}pez, C.~E., \& Castillo, D.~J.\ 2007, AJ, 134, 195 

\bibitem[]{} Cassisi, S., Salaris, M., \& Irwin, A.W. 2003, ApJ, 588, 862

\bibitem[Castelli 
\& Kurucz(2004)]{2004astro.ph..5087C} Castelli, F., \& Kurucz, R.~L.\ 2004, arXiv:astro-ph/0405087 

\bibitem[Castellani \& Castellani(1993)]{1993ApJ...407..649C} Castellani, M., \& Castellani, V.\ 1993, ApJ, 407, 649 

\bibitem[Castellani et 
al.(1994)]{1994A&A...282..771C} Castellani, M., Castellani, V., Pulone, L., \& Tornambe, A.\ 1994, A\&A, 282, 771 

\bibitem[Castellani et 
al.(2006)]{2006A&A...446..569C} Castellani, V., Iannicola, G., Bono, G., Zoccali, M., Cassisi, S., \& Buonanno, R.\ 2006, A\&A, 446, 569 

\bibitem[Cohen(1978)]{1978ApJ...223..487C} Cohen, J.~G.\ 1978, ApJ, 223, 
487 

\bibitem[Corwin et 
al.(2004)]{2004A&A...421..667C} Corwin, T.~M., Catelan, M., Borissova, J., \& Smith, H.~A.\ 2004, A\&A, 421, 667 

\bibitem[Da Costa et al.(2009)]{2009ApJ...705.1481D} Da Costa, G.~S., Held, 
E.~V., Saviane, I., \& Gullieuszik, M.\ 2009, ApJ, 705, 1481

\bibitem[Dalessandro et al.(2008)]{2008ApJ...677.1069D} Dalessandro, E., 
Lanzoni, B., Ferraro, F.~R., Rood, R.~T., Milone, A., Piotto, G., 
\& Valenti, E.\ 2008, ApJ, 677, 1069 

\bibitem[Dalessandro et al.(2009)]{2009ApJS..182..509D} Dalessandro, E., 
Beccari, G., Lanzoni, B., Ferraro, F.~R., Schiavon, R., 
\& Rood, R.~T.\ 2009, ApJS, 182, 509 

\bibitem[D'Antona et 
al.(2002)]{2002A&A...395...69D} D'Antona, F., Caloi, V., Montalb{\'a}n, J., Ventura, P., \& Gratton, R.\ 2002, A\&A, 395, 69

\bibitem[D'Antona 
\& Caloi(2004)]{2004ApJ...611..871D} D'Antona, F., \& Caloi, V.\ 2004, ApJ, 611, 871 

\bibitem[D'Antona et al.(2005)]{2005ApJ...631..868D} D'Antona, F., 
Bellazzini, M., Caloi, V., Pecci, F.~F., Galleti, S., 
\& Rood, R.~T.\ 2005, ApJ, 631, 868 


\bibitem[D'Cruz et al.(1996)]{1996ApJ...466..359D} D'Cruz, N.~L., Dorman, 
B., Rood, R.~T., \& O'Connell, R.~W.\ 1996, ApJ, 466, 359 

\bibitem[Decressin et 
al.(2007)]{2007A&A...464.1029D} Decressin, T., Meynet, G., Charbonnel, C., Prantzos, N., \& Ekstr{\"o}m, S.\ 2007, A\&A, 464, 1029 

\bibitem[Decressin et 
al.(2008)]{2008A&A...492..101D} Decressin, T., Baumgardt, H., \& Kroupa, P.\ 2008, A\&A, 492, 101
 
\bibitem[Dorman et al.(1993)]{1993ApJ...419..596D} Dorman, B., Rood, R.~T., 
\& O'Connell, R.~W.\ 1993, ApJ, 419, 596 

\bibitem[D'Ercole et al.(2008)]{2008MNRAS.391..825D} D'Ercole, A., 
Vesperini, E., D'Antona, F., McMillan, S.~L.~W., 
\& Recchi, S.\ 2008, MNRAS, 391, 825 



\bibitem[Dieball et al.(2005)]{2005ApJ...625..156D} Dieball, A., Knigge, 
C., Zurek, D.~R., Shara, M.~M., \& Long, K.~S.\ 2005, ApJ, 625, 156 


\bibitem[Dorman et al.(1993)]{1993ApJ...419..596D} Dorman, B., Rood, R.~T., 
\& O'Connell, R.~W.\ 1993, ApJ, 419, 596 

\bibitem[Dotter(2008)]{2008ApJ...687L..21D} Dotter, A.\ 2008, ApJl, 687, 
L21

\bibitem[Dotter et al.(2010)]{2010ApJ...708..698D} Dotter, A., et al.\ 
2010, ApJ, 708, 698 

\bibitem[Fabbri et al. (2010)]{in preparation} Fabbri et al. 2010, in preparation

\bibitem[Ferraro et 
al.(1990)]{1990A&AS...84...59F} Ferraro, F.~R., Clementini, G., \
 Fusi Pecci, F., Buonanno, R., \& Alcaino, G.\ 1990, A\&AS, 84, 59

\bibitem[Ferraro et al.(1998)]{1998ApJ...500..311F} Ferraro, F.~R., 
Paltrinieri, B., Pecci, F.~F., Rood, R.~T., 
\& Dorman, B.\ 1998, ApJ, 500, 311 

\bibitem[Ferraro et al.(1999)]{1999AJ....118.1738F} Ferraro, F.~R., 
Messineo, M., Fusi Pecci, F., de Palo, M.~A., Straniero, O., Chieffi, A., 
\& Limongi, M.\ 1999, AJ, 118, 1738 

\bibitem[Ferraro et al.(2001)]{2001ApJ...561..337F} Ferraro, F.~R., 
D'Amico, N., Possenti, A., Mignani, R.~P., 
\& Paltrinieri, B.\ 2001, ApJ, 561, 337 


\bibitem[Ferraro et al.(2003)]{2003ApJ...596L.211F} Ferraro, F.~R., 
Possenti, A., Sabbi, E., \& D'Amico, N.\ 2003, ApJl, 596, L211 

\bibitem[Ferraro et al.(2004)]{2004ApJ...603L..81F} Ferraro, F.~R., 
Sollima, A., Pancino, E., Bellazzini, M., Straniero, O., Origlia, L., 
\& Cool, A.~M.\ 2004, ApJl, 603, L81 

\bibitem[Ferraro et al.(2009)]{2009Natur.462..483F} Ferraro, F.~R., et al.\ 
2009, Nature, 462, 483 

\bibitem[Girardi et 
al.(2002)]{2002A&A...391..195G} Girardi, L., Bertelli, G., Bressan, A., Chiosi, C., Groenewegen, M.~A.~T., Marigo, P.,
Salasnich, B., \& Weiss, A.\ 2002, A\&A, 391, 195

\bibitem[Gratton et 
al.(2000)]{2000A&A...358..671G} Gratton, R.~G., Carretta, E., Matteucci, F., \& Sneden, C.\ 2000, A\&A, 358, 671 

\bibitem[Gratton et al.(2010)]{2010arXiv1004.3862G} Gratton, R.~G., 
Carretta, E., Bragaglia, A., Lucatello, S., 
\& D'Orazi, V.\ 2010, arXiv:1004.3862


\bibitem[Harris(1974)]{1974ApJ...192L.161H} Harris, W.~E.\ 1974, ApJl, 
192, L161 

\bibitem[Harris (1996)]{har96} Harris, W.E. 1996, AJ, 112, 1487

\bibitem[Holtzman et al.(1995)]{1995PASP..107.1065H} Holtzman, J.~A., 
Burrows, C.~J., Casertano, S., Hester, J.~J., Trauger, J.~T., Watson, 
A.~M., \& Worthey, G.\ 1995, PASP, 107, 1065 

\bibitem[Hui-Bon-Hoa et al.(2000)]{2000ApJ...535L..43H} Hui-Bon-Hoa, A., 
LeBlanc, F., \& Hauschildt, P.~H.\ 2000, ApJl, 535, L43 

\bibitem[Iannicola et al.(2009)]{2009ApJ...696L.120I} Iannicola, G., et 
al.\ 2009, ApJl, 696, L120

\bibitem[Lanzoni et al.(2007)]{2007ApJ...668L.139L} Lanzoni, B., 
Dalessandro, E., Ferraro, F.~R., Miocchi, P., Valenti, E., 
\& Rood, R.~T.\ 2007, ApJl, 668, L139 

\bibitem[Lanzoni et al.(2010)]{2010arXiv1005.2847L} Lanzoni, B., et al.\ 
2010, arXiv:1005.2847

\bibitem[Lee et al.(1994)]{1994ApJ...423..248L} Lee, Y.-W., Demarque, P., 
\& Zinn, R.\ 1994, ApJ, 423, 248 


\bibitem[Mackey \& van den Bergh(2005)]{2005MNRAS.360..631M} Mackey, A.~D., 
\& van den Bergh, S.\ 2005, MNRAS, 360, 631 

\bibitem[Marino et 
al.(2009)]{2009A&A...505.1099M} Marino, A.~F., Milone, A.~P., Piotto, G., Villanova, S., Bedin, L.~R., Bellini, A., \& Renzini, A.\ 2009, A\&A, 505, 1099 

\bibitem[Michaud et al.(2008)]{2008ApJ...675.1223M} Michaud, G., Richer, 
J., \& Richard, O.\ 2008, ApJ, 675, 1223 

\bibitem[Miller Bertolami et 
al.(2008)]{2008A&A...491..253M} Miller Bertolami, M.~M., Althaus, L.~G., Unglaub, K., \& Weiss, A.\ 2008, A\&A, 491, 253 

\bibitem[Moehler et al. (2004)]{moehler2808} Moehler, S., Sweigart, A.~V.,
Landsman, W.~B., Hammer, N.~J., \& Dreizler, S.\ 2004, A\&A, 415, 313

\bibitem[Moni Bidin et 
al.(2006)]{2006A&A...451..499M} Moni Bidin, C., Moehler, S., Piotto, G., Recio-Blanco, A., Momany, Y., \& M{\'e}ndez,
R.~A.\ 2006, A\&A, 451, 499 

\bibitem[Montegriffo et al.(1995)]{1995MNRAS.276..739M} Montegriffo, P., 
Ferraro, F.~R., Fusi Pecci, F., \& Origlia, L.\ 1995, MNRAS, 276, 739 

\bibitem[Pace et 
al.(2006)]{2006A&A...452..493P} Pace, G., Recio-Blanco, A., Piotto, G., \& Momany, Y.\ 2006, A\&A, 452, 493

\bibitem[Pancino et al.(2002)]{2002ApJ...568L.101P} Pancino, E., Pasquini, 
L., Hill, V., Ferraro, F.~R., \& Bellazzini, M.\ 2002, ApJl, 568, L101

\bibitem[Pietrinferni et al.(2006)]{2006ApJ...642..797P} Pietrinferni, A., 
Cassisi, S., Salaris, M., \& Castelli, F.\ 2006, ApJ, 642, 797 

\bibitem[Piotto et al.(2005)]{2005ApJ...621..777P} Piotto, G., et al.\ 
2005, ApJ, 621, 777 

\bibitem[]{} Piotto, G., Bedin, L. R., Anderson, J., King, I. R., Cassisi, S., Milone, A. P., Villanova, S., Pietrinferni, A., \& Renzini, A. 2007, ApJ, 661, L53

\bibitem[]{} Quievy, D., Charbonneau, P., Michaud, G., \& Richer,
  J. 2009, A\&A, 500, 1163

\bibitem[Recio-Blanco et 
al.(2006)]{2006A&A...452..875R} Recio-Blanco, A., Aparicio, A., Piotto, G., de Angeli, F., \& Djorgovski, S.~G.\ 2006, A\&A, 452, 875 

\bibitem[Rich et al.(1997)]{1997ApJ...484L..25R} Rich, R.~M., et al.\ 1997,  ApJL, 484, L25

\bibitem[Rood(1973)]{1973ApJ...184..815R} Rood, R.~T.\ 1973, ApJ, 184, 815 

\bibitem[Rood et al.(1993)]{1993ASPC...48..218R} Rood, R.~T., Crocker, 
D.~A., Fusi Pecci, F., Ferraro, F.~R., Clementini, G., 
\& Buonanno, R.\ 1993, The Globular Cluster-Galaxy Connection, 48, 218 

\bibitem[Rood et al.(2008)]{2008MmSAI..79..383R} Rood, R.~T., Beccari, G., 
Lanzoni, B., Ferraro, F.~R., Dalessandro, E., 
\& Schiavon, R.~P.\ 2008, Memorie della Societa Astronomica Italiana, 79, 383

 
\bibitem[Rutledge et al.(1997)]{1997PASP..109..883R} Rutledge, G.~A., 
Hesser, J.~E., Stetson, P.~B., Mateo, M., Simard, L., Bolte, M., Friel, 
E.~D., \& Copin, Y.\ 1997, PASP, 109, 883 
 

\bibitem[Samus et al.(2009)]{2009PASP..121.1378S} Samus, N.~N., Kazarovets, 
E.~V., Pastukhova, E.~N., Tsvetkova, T.~M., 
\& Durlevich, O.~V.\ 2009, PASP, 121, 1378 


\bibitem[Sollima et al.(2005)]{2005MNRAS.357..265S} Sollima, A., Ferraro, 
F.~R., Pancino, E., \& Bellazzini, M.\ 2005, MNRAS, 357, 265 

\bibitem[Sosin et al.(1997)]{1997ApJ...480L..35S} Sosin, C., et al.\ 1997, 
ApJl, 480, L35 

\bibitem[Ventura 
\& D'Antona(2009)]{2009A&A...499..835V} Ventura, P., \& D'Antona, F.\ 2009, A\&A, 499, 835 

\bibitem[Villanova et al.(2007)]{2007ApJ...663..296V} Villanova, S., et 
al.\ 2007, ApJ, 663, 296 

\bibitem[Zinn 
\& West(1984)]{1984ApJS...55...45Z} Zinn, R., \& West, M.~J.\ 1984, ApJS, 55, 45 


\end{thebibliography}
\end{document}